\documentclass{emulateapj}

\newcommand     \etal    {et al.}
\newcommand     \ergss  {ergs s$^{-1}$}
\newcommand     \ans     {\rm \AA}
\newcommand	\kms    {\ensuremath{\mathrm{km~s}^{-1}}} 
\newcommand	\msun {\ensuremath{M_{\sun}}}
\newcommand	\msunyr {\ensuremath{M_{\sun}~\mathrm{yr}^{-1}}}
\shortauthors{Bai \etal}
\begin{document}
\title{The First Mid-IR View of the Star-forming Properties 
       of Nearby Galaxy Groups}
\author{Lei\,Bai\altaffilmark{1},
Jesper\,Rasmussen\altaffilmark{1,4},
John S.\,Mulchaey\altaffilmark{1},
Ali\,Dariush\altaffilmark{2},
Somak\,Raychaudhury\altaffilmark{3},
Trevor J.\, Ponman\altaffilmark{3}
}
\email{leibai@obs.carnegiescience.edu}
\altaffiltext{1}{The Observatories of the Carnegie Institution of Washington, 813 Santa Barbara Street, Pasadena, 
CA 91101, USA}
\altaffiltext{2}{Cardiff School of Physics and Astronomy, Cardiff University, Queens Buildings, The Parade, Cardiff, CF24 3AA, UK}
\altaffiltext{3}{School of Physics and Astronomy, University of Birmingham, 
Birmingham B15 2TT, UK}
\altaffiltext{4}{$Chandra$ Fellow}

\begin{abstract}
  We present the first mid-IR study of galaxy groups in the nearby
  Universe based on {\em Spitzer} MIPS observations of a sample of
  nine redshift-selected groups from the XMM-IMACS (XI) project, at
  $z=0.06$.  We find that on average the star-forming (SF) galaxy
  fraction
  in the groups is about 30\% lower than the value in the field and
  30\% higher than in clusters.  The SF fractions do not show any
  systematic dependence on group velocity dispersion, total stellar
  mass, or the presence of an X-ray emitting intragroup medium, but a
  weak anti-correlation is seen between SF fraction and projected
  galaxy density.  However, even in the densest regions, the SF
  fraction in groups is still higher than that in cluster outskirts,
  suggesting that preprocessing of galaxies in group environments is
  not sufficient to explain the much lower SF fraction in clusters.
  The typical specific star formation rates (SFR/$M_\ast$) of SF
  galaxies in groups are similar to those in the field across a wide
  range of stellar mass ($M_\ast>10^{9.6}\msun$), favoring a quickly
  acting mechanism that suppresses star formation to explain the
  overall smaller fraction of SF galaxies in groups.
  If galaxy-galaxy interactions are responsible, then the extremely
  low starburst galaxy fraction ($<1\%$) implies a short timescale
  ($\sim0.1$ Gyr) for any merger-induced starburst stage. Comparison
  to two rich clusters shows that clusters contain a population of massive
  SF galaxies with very low SFR (14\% of all the galaxies with $M_\ast>10^{10}\,\msun$), possibly as a consequence of ram pressure
  stripping being less efficient in removing gas from more massive
  galaxies.
\end{abstract}


\section{INTRODUCTION}
The local galaxy population presents a clear bimodality in many
different properties: blue galaxies with active star formation and
late-type morphologies vs red, quiescent galaxies with early-type
morphologies \citep[e.g.,][]{Strateva01,Baldry04,Balogh04}.  This
bimodality is ubiquitous, extending from galaxy clusters to groups and
to the general field \citep{Lewis02,Gomez03}. The physical origin of
this bimodality remains one of the most puzzling questions of galaxy
formation and evolution.  Is the difference of the two distinct
populations seeded in the early stages of galaxy formation (the
so-called \lq nature\rq\ scenario), or is it the end result of a
transformation driven by environment (the \lq nurture\rq\ scenario)?
Strong evidence favoring the nurture scenario comes from the drastic
change of the fraction of galaxies in these two populations in
different environments: the fraction of passive galaxies increases
with increasing galaxy density.
However, such a correlation alone does not directly imply a nurture
scenario.  The fraction of galaxies in different populations also
depends on galaxy stellar mass \citep{Kauffmann03}, which could have a
different distribution in low- and high-density regions seeded at the
time of galaxy formation.  Hence, to fully understand galaxy
evolution, we need to disentangle the stellar-mass and environment
dependence \citep[e.g.,][]{Baldry06,Iovino09,Kovac09} and identify the mechanisms
responsible for establishing the bimodality in galaxy properties.

Galaxy groups, as the most common galaxy associations, contain about
50\% of the galaxy population at the present day \citep{Geller83,
Tully88, Eke04, Eke05}.  The characteristic depth of the potential
wells of groups is similar to those of individual galaxies, and the
velocities of galaxies within groups are only a few hundred \kms,
comparable to the internal velocity of galaxies.  Under these
circumstances, galaxies interact strongly with one another, and with
the group as a whole.  Such interactions could transform the
morphology of galaxies, induce starbursts, and thereby turn active,
late-type galaxies into quiescent, early-type galaxies.  In addition,
the group environment could also transform galaxies via ram pressure
stripping of their hot gas halos, eventually suffocating star
formation \citep{Rasmussen06a,Kawata08,McCarthy08}. Hence, not only
are groups the most common environmental phase experienced by galaxies
during their evolution, they also have the potential to strongly
affect large populations of galaxies and thereby help to explain the
ubiquitous bimodality in galaxy properties.

Another important implication of galaxy transformations in groups is
the pre-processing of galaxies before they fall into clusters
\citep{Zabludoff98}. In a hierarchical structure formation scenario,
clusters are built up by the accretion of smaller structures, e.g.,
isolated galaxies and groups.  However, it is still under debate if
the majority of cluster galaxies were ever located in groups before
being acquired by clusters \citep{Berrier09, McGee09}.  If the
fraction of cluster galaxies accreted in groups is substantial,
preprocessing of galaxy properties by the group environment could play
a major role in forming the predominantly passive population in
clusters. Additional physical mechanisms which only work efficiently
in the cluster environment, such as galaxy harassment and ram pressure
stripping of cold galactic gas, could further affect galaxy properties
but would possibly only be of secondary importance.

Despite the importance of the group environment for the global galaxy
population, our understanding of groups is still very limited.  There
have been many studies of group galaxies trying to address these
questions.  Due to the low galaxy density contrast against the field,
many previous group studies focused on X-ray luminous groups, which
are mostly virialized groups \citep{Mulchaey98,Zabludoff98}.  However,
when referring to the majority of galaxies as being located in groups,
this applies specifically to {\it optically} selected groups,
typically identified through a \lq\lq friends-of-friends\rq\rq \
analysis of galaxy redshift survey data \citep{Eke04,Balogh04}.  Such
groups span a wide range of evolutionary states, including systems in
the process of collapsing (like the Local Group), systems in the
throes of strong galaxy interactions or subgroup mergers, and fully
virialized systems.
To fully assess the importance of the group environment on galaxy
evolution, complete samples of galaxy groups extending to poor systems
are essential.  There have also been studies of optically selected
groups based on large-area galaxy surveys \citep{Balogh04,Weinmann06}.
However, due to the limited survey depth, many of the poor groups in
those studies only contain a handful of known group members.  To fully
understand the dynamics and galaxy content of each group, we need to
probe to significantly fainter optical magnitudes than is usually
possible with such large surveys.

For this purpose, we have started the XMM-IMACS (XI) Groups Project
(\citealt{Rasmussen06b}; hereafter Paper~I) to carry out a
multi-wavelength study of a statistically representative nearby group
sample.  We selected 25 groups with velocity dispersion $\sigma <500$
\kms\ from the group catalog of \citet{Merchan02}, which was carefully
derived from a friends-of-friends analysis of the 2dF redshift survey
\citep{Colless01}.  The groups are selected in a narrow redshift range
($0.06<z<0.062$) to minimize the redshift dependence of group
properties.  To fully represent the poor group population, the XI
group sample was selected to span a wide range in group properties,
e.g., velocity dispersion, virial radius, and number of group members
\citep{Rasmussen06b}.  To extract membership information down to faint
magnitudes, multislit spectroscopy of these groups has been performed
using IMACS on the 6.5-m Baade/Magellan telescope \citep{Bigelow03},
and for nine of the groups we have also obtained X-ray observations
using {\em XMM-Newton}. At the chosen group redshifts, the fields of
view of both IMACS and {\em XMM-Newton} cover the typical virial
radius of the groups, $\sim$~1~Mpc.

To assess the star formation properties of galaxies in these groups,
we carried out {\em Spitzer} MIPS 24~\micron\ imaging of the XI group
sample.  Different from emission line or UV luminosity often used in
previous group studies to measure the star formation rate (SFR) of
galaxies, the mid-IR emission from the dust heated by a young stellar
population is a robust star formation indicator unaffected
by extinction.  Thus, IR emission is not only critical in giving us an
unbiased view of the star formation properties of the group
population, but more importantly, is essential to identify dusty
starbursts in which the majority of the star formation is obscured and
not seen in optical spectra \citep{Liu95}. Although there have been
many studies of star-forming (SF) galaxies using IR observations in
clusters
\citep{Fadda08,Geach06,Marcillac07,Bai06,Bai07,Bai09,Dressler09,Saintonge08,Haines09a,Haines09b,Wolf09,Mahajan10},
such studies of groups have been lacking. While \citet{Tran09}
detected an excess of 24 \micron\ SF galaxies in groups compared to
the field, we note that their sample represents a somewhat peculiar
case of a super-group in the process of forming a massive cluster.

In this paper, we present the first IR study of star formation in a
representative, nearby sample of groups, based on nine groups in our
XI sample.  These are the first nine groups for which we have complete
spectroscopic data and X-ray observations.  We analyze the star
forming properties of group galaxies from their 24~\micron\ emission
and compare them to those of field and cluster galaxies. In
Section~\ref{sec,obs}, we describe the observations and data
reduction, and in Section~\ref{sec,general} discuss the general
properties of the groups.  In Section~\ref{sec,results}, we present
the SF galaxy fractions and specific SFRs of group galaxies, and
compare them to results for galaxies in clusters and the field.  The
results are discussed in Section~\ref{sec,discuss} and summarized in
Section~\ref{sec,summary}. Throughout this paper, we assume a
$\Lambda$CDM cosmology with parameter set $(h,\Omega_{0},\Lambda_{0})
= (0.7,0.3,0.7)$.

\section{OBSERVATIONS AND DATA}\label{sec,obs}
\subsection{XI Groups}
\subsubsection{Optical Imaging and Spectroscopy}
Optical images of the groups were taken in the Bessel {\em BVR}
filters with the Wide Field Reimaging CCD (WFCCD) on the 100-inch du
Pont telescope at Las Campanas.  Source extraction and photometry were
done with {\sc SExtractor} \citep{Bertin96}.  The magnitudes were
corrected for Galactic extinction using the dust map from
\citet{Schlegel98}.  Follow-up multi-object spectroscopy of galaxies
in the group fields was obtained with the IMACS spectrograph on the
Baade/Magellan telescope.  The spectra were taken with the
300~lines~mm$^{-1}$ prism on the f/2 camera, which covers a wavelength
range of 3900--10000~\ans\ at a dispersion of 1.34~\ans~pixel$^{-1}$.
Spectroscopic targets were selected based on $R$-band magnitudes,
prioritizing the brighter sources. The spectra were reduced using the
COSMOS software package.  We refer readers to Paper~I for more details
on the target selection, observing strategy, and data reduction.  The
redshifts of the galaxies were determined by cross-correlating with
SDSS galaxy templates. Typical errors in the redshift measurements
were $\sim 50$~\kms.

In most group fields, our optical imaging and spectroscopy only cover
the central $20\arcmin\times20\arcmin$ region, which corresponds to
$1.4 \times 1.4$~Mpc$^{2}$ at the relevant redshifts. To extend the
spatial coverage to larger radii and better match the larger extent of
the IR data, our spectroscopic data were complemented with redshift
measurements from NED and the 6dF Galaxy Survey
\citep[6dFGS,][]{Jones09}. Most of the redshift measurements in NED
are from 2dFGRS, which probes down to a magnitude limit of $b_{J}\leq
19.45$, while the 6dFGS extends down to $b_{J}\leq 16.75$.  Some of
those galaxies are outside the region covered by our imaging data. For
these galaxies, we obtained $R$-band magnitudes from the SuperCOSMOS
Sky Surveys \citep[SSS,][]{Hambly01}, which are, where they overlap,
consistent with the photometry derived from our imaging data. Our
final spectroscopic catalogs are $>80\%$ complete down to $R=18$ in
the central $<10\arcmin$ (0.7 Mpc) region and $>50\%$ complete out to
a radius of $25\arcmin$ (1.7 Mpc).

\subsubsection{X-ray Observations}
All nine groups discussed in this paper were observed by {\em
XMM-Newton}. The details of the X-ray observations and analysis can be
found in Paper~I and \citet{Shen07}. The nominal exposure time in all
fields is $\sim 20$~ks, although the useful exposure time varies due
to the presence of background flares in several of the
observations. The typical detection limit for X-ray point sources is
$\sim 2\times10^{40}$~erg~s$^{-1}$ (0.3--2~keV), assuming a power-law
spectrum of photon index of 1.7 subject to Galactic absorption.
Extended emission from a hot intragroup medium is clearly detected in
three groups, MZ\,10451, MZ\,4577, and MZ\,9014.  MZ\,10451 is the
most X-ray luminous group in our sample and has $L_{X}\sim
2\times10^{42}$~ergs~s$^{-1}$ \citep{Rasmussen10}, comparable to
typical X-ray selected groups with similar velocity dispersion. Both
MZ\,4577 and MZ\,9014 have low levels of extended emission
\citep[$L_{X}\sim10^{41}$ ergs s$^{-1}$,][]{Rasmussen06b}.  A more
detailed analysis of the {\em XMM-Newton} data will be presented in a
future paper.

\subsubsection{MIPS 24~\micron\ Imaging and SFR Estimation}\label{sec:SFR}
The 24~\micron\ images of the groups were taken by MIPS
\citep{Rieke04} on {\em Spitzer} in the medium scan map mode during
2007--2008. The observed field for each group was a rectangular region
about $20\arcmin \times 45\arcmin$ ($1.4 \times 3.1$ Mpc$^{2}$) in
size centered at the group center from the \citet[hereafter
MZ]{Merchan02} catalog.  The data were processed with the MIPS Data
Analysis Tool \citep[DAT version 3.02;][]{Gordon05}, and
array-averaged background subtraction was applied to improve the
signal-to-noise ratio.  The final mosaics have an exposure time of
$\sim80$~sec~pixel$^{-1}$ and a spatial resolution of
$\sim5\arcsec$. We used {\sc SExtractor} to extract sources and
measure their 24~\micron\ flux within an adaptive Kron aperture ({\tt
flux\_auto}). The sensitivity of the data varies slightly from field
to field depending on the IR background level, but the 3-$\sigma$
point source detection limit is $\leq$~0.35~mJy in all fields.

We correlated galaxies with the nearest 24 \micron\ sources projected
within $5\arcsec$.  This matching radius takes into consideration the
possible physical displacement between the optical and 24 \micron\
brightness centroids.  To determine the SFR from the 24~\micron\ flux,
we use the conversion given by \citet{Rieke09}.  However, the IR luminosity limit of our data, log$\,L_{\rm
IR}/L_{\sun}=8.9$, is fainter than the lower limit to which this
formula is applicable (log$\,L_{\rm IR}/L_{\sun}=9.7$).  For galaxies
with lower IR luminosity, the ratio of escaping UV photons to
UV photons absorbed by dust increases \citep{Buat07}. At log$\,L_{\rm
IR}/L_{\sun}\approx 11$, the average leakage is only about 2.5\%, and
this is the correction factor adopted by \citet{Rieke09} in the
derivation of their conversion formula.  However, for less luminous
galaxies, this correction is insufficient, as the average leakage
increases to about 50\% at log$\,L_{\rm IR}/L_{\sun}=8.5$
\citep{Buat07}. To extend the SFR conversion to less luminous galaxies
(log$\,L_{\rm IR}/L_{\sun}<11$), instead of using an average leakage
fraction of 2.5\%, we estimate the leakage as a function of $L_{\rm
IR}$ using the average $L_{\rm IR}/L_{\rm UV}$ vs. $L_{\rm bol}$
relation provided by \citet[their figure 7]{Buat07}, where $L_{\rm
bol}=L_{\rm UV}+0.7L_{\rm IR}$. We then correct the SFR derived from
Rieke's formula with this leakage fraction.  This correction is
generally small, increasing the derived SFR at most by a factor of 2,
and it only affects galaxies with low SFRs.  With this conversion, the
24~\micron\ detection limit of our observations translates into an SFR
limit of 0.1~\msunyr.

In this work, we assume that the 24~\micron\ emission of galaxies is
predominantly from dust heated by young stars. This is true for
star-forming galaxies and H{\sc ii} regions, based on which the
24~\micron\ to SFR conversion is calibrated \citep[e.g.,][]{Rieke09}.
However, for galaxies with small 24~\micron\ luminosities, especially
early-type galaxies, the contribution from cold dust heated by an
evolved stellar population becomes significant.  \citet{Temi09a,
Temi09b} have shown that for elliptical galaxies there is a
correlation between 24~\micron\ luminosity and the near-IR luminosity
$L_{K_{s}}$, consistent with the expectation of mid-IR emission
originating from cold dust surrounding the old, mass-losing red-giant
stars.  With this correlation, we can estimate the cold dust
contribution to the 24~\micron\ emission from $L_{K_{s}}$.  To do so,
we retrieved $K_{s}$ band luminosities of our group galaxies from the
Two Micron All Sky Survey (2MASS), which is complete down to $R\sim16$
at the group redshift.  For group galaxies too faint to be detected by
2MASS, we estimate their $K_{s}$ luminosity from $R$ magnitudes.  This
introduces an extra uncertainty in our estimates, but for these less
massive galaxies, the cold dust contribution is generally minimal
compared to the 24~\micron\ detection limit to which we are sensitive.
Following \citet{Temi09a}, we regard the 24~\micron\ emission of all
galaxies with log\,$(L_{24}/L_{K_{s}}) < 30.5$ (with $L_{24}$ in
erg~s$^{-1}$ and $L_{K_{s}}$ in $L_\odot$) as coming exclusively from
cold dust, setting their SFR to zero; for galaxies above this limit,
we subtract a cold dust contribution, log\,$L_{\rm
24,cold}=(1.01\pm0.05)$\,log\,$L_{K_{s}}+30.1\pm0.5$, from the
observed $L_{24}$ before converting the latter into an SFR estimate.
Overall, this cold dust correction for our sample is very small: for
the 24~\micron\ detected galaxies in our groups, none has $L_{24}$
consistent with being from cold dust alone, and the majority has a
cold dust contribution less than 10\%.

Another complication of using mid-IR emission to derive SFR is the
contamination from AGN activity.  In this case, the dust is, at least
partly, heated by AGNs and the correlation between SFRs and
24~\micron\ flux no longer holds.  To identify AGN, we cross-match
group members with the X-ray point sources detected in each field.
Using a matching radius of 10$\arcsec$, we found unambiguous X-ray
counterparts, with $L_{X}>10^{41}$~ergs~s$^{-1}$, for five galaxies in
the nine groups.  If we only consider the galaxies with $M_{R}<-20$
within the {\em XMM} field of view, the fraction of group galaxies
with $L_{X}>10^{41}$~ergs~s$^{-1}$ is $6^{+12}_{-6}$\%, consistent
with the 5\% found in clusters \citep{Martini06}.  Of those five X-ray
bright galaxies, only two are also detected at 24~\micron. This only
accounts for a small fraction ($\sim2\%$) of the IR-detected galaxies.
This fraction is also consistent with the result found in the A901/902
supercluster \citep{Gallazzi09}.  In addition to X-ray luminous AGN,
\citet{Shen07} also identified five optical AGN with no X-ray
counterpart within our groups. All of these are bright at 24~\micron.
Even though these galaxies show emission line ratios typical of AGN,
their IR emission is probably still dominated by star formation
\citep[e.g.,][]{Brand09}.

Because of the difficulty in removing the AGN contribution to the IR
emission, and the small number of group galaxies with known AGN
signatures, we do not exclude these AGN in our IR-selected SF galaxy
sample.  In the following sections, we compare our group SF galaxies
to field and cluster SF galaxies and we note that these comparison
samples of IR-selected SF galaxies may also be contaminated by AGN.
However, the inferred dependence of AGN fraction on environment is
generally weak \citep{Kauffmann04,Martini06}. In particular, the
results of \citet{Shen07} show that, down to the limiting magnitude of
$M_R=-20$ that we consider for SF galaxies in
Sections~\ref{sec,results} and \ref{sec,discuss}, the overall AGN
fraction within our groups is consistent with that of rich
clusters. Hence, we do not expect environmental differences in AGN
activity to have a significant impact on our results.

\subsection{Field and Cluster Comparison Samples}
We further compiled a field galaxy sample from 23 XI group fields with
MIPS data.  For each field, we retrieved all the galaxies with
redshifts from NED, where most of them are from the 2dFGRS survey.
The redshift histogram of each field was examined to exclude
foreground/background clusters and groups, resulting in a field sample
of 77 galaxies in the redshift range of $0.02 < z < 0.57$.  The
$R$-band magnitudes of these galaxies were obtained from the 2dFGRS
photometric catalog.  Among these field galaxies, 45 are brighter than
$M_{R}=-19$.

For comparison, we also compiled a cluster galaxy sample from two
local rich clusters: the Coma cluster ($z=0.023$) and the Abell 3266
cluster (A3266, $z=0.06$).  Both clusters have been observed with MIPS
at 24 \micron\ to similar depth as the XI groups.  More details on the
MIPS data of these two clusters are reported in \citet{Bai06, Bai09}.
Altogether, we have around 600 cluster galaxies with $M_{R}<-19$.  We
updated the 24 \micron\ SFR conversion for these galaxies with the new
method described in Section~\ref{sec:SFR}. Compared to the group and
field galaxies, the cold dust correction is more important for cluster
galaxies.  About 20\% of the cluster members have a cold dust
contribution of more than 20\%, and about 10\% have a 24 \micron\ flux
consistent with being from cold dust alone.

\section{General Properties of the Groups}\label{sec,general}
\subsection{Group Member Selection}
To select group members, we performed an iterative 3-$\sigma$ clipping
in redshift space using the bi-weight mean and dispersion
\citep{Beers90} until the number of group members converged. In most
groups, all galaxies with velocity $\pm2000$~km~s$^{-1}$ within the
group mean and within 25$\arcmin$ ($\sim$~1.7~Mpc) of the group
centers were included in the calculation.  For MZ\,3849 and MZ\,9307,
narrower velocity ranges were employed (1300 and 1500~km~s$^{-1}$,
respectively), in order to exclude contamination from nearby unrelated
structures. In all cases, the calculations converged in one or two
iterations.  In total, we identified 273 group members in the nine
groups. With the new membership lists, we updated the group redshifts
and velocity dispersions, as listed in Table~\ref{tab_1}.  The
1-$\sigma$ errors on the velocity dispersions were derived from
bootstrap calculations.  The velocity dispersions of our group sample
range from $\sim 100$--500~\kms.  We also calculated the velocity
dispersions using only galaxies within 1~Mpc from the group centers,
confirming that the differences in resulting velocity dispersions are
well within the estimated errors.

\begin{figure}
\epsscale{0.8}
\plotone{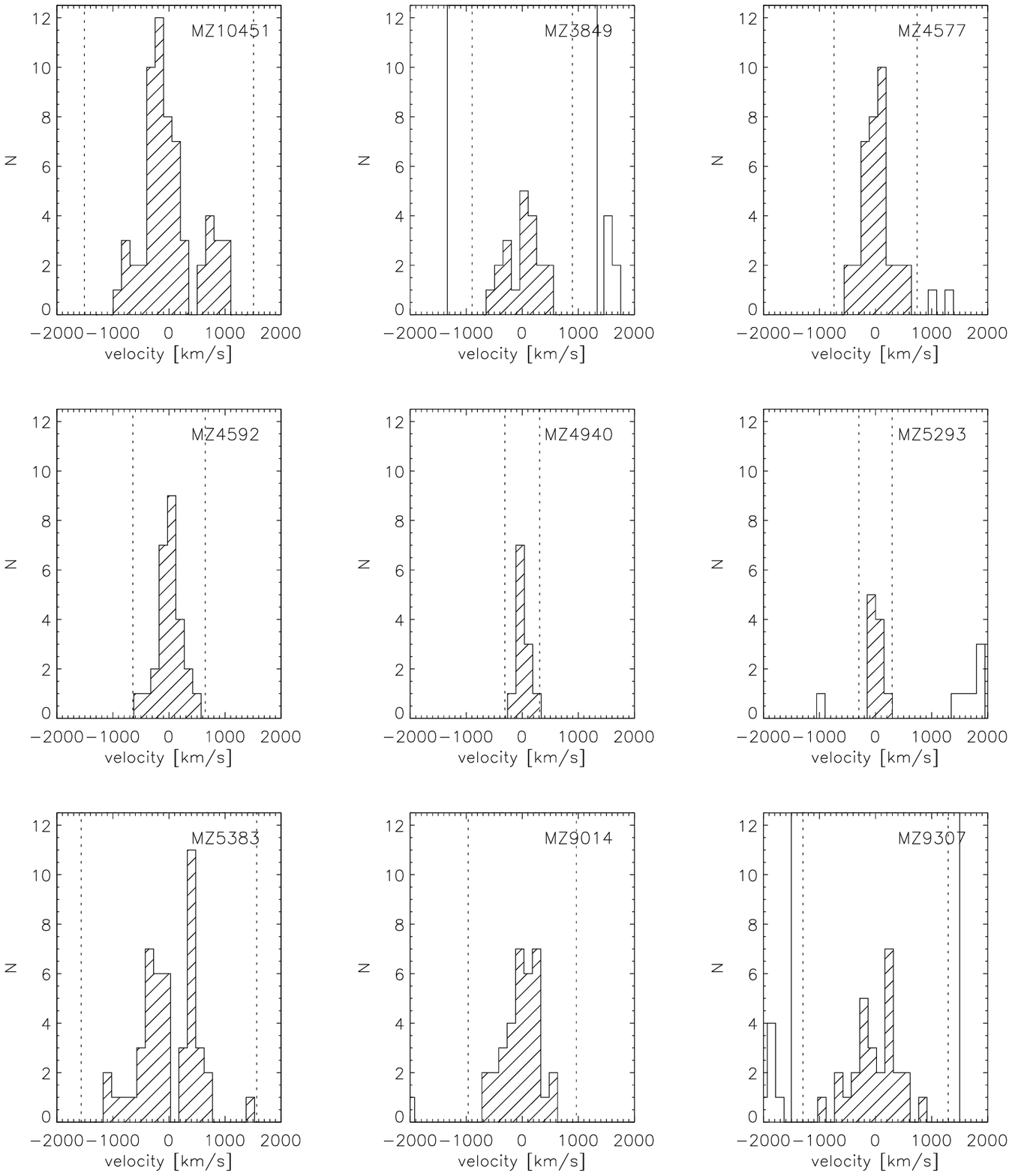}
\caption{Galaxy velocity histograms of the groups. Shaded regions
  represent identified group members. Dotted vertical lines indicate
  the $\pm3\sigma$ range. Solid vertical lines for MZ\,3849 and
  MZ\,9307 show the imposed velocity cuts needed to exclude nearby
  contamination. In other groups, this velocity cut is
  $\pm2000$~\kms.}
\label{f_velhist}
\end{figure}

The galaxy velocity histogram in each field is plotted in
Figure~\ref{f_velhist} in bins of 150~\kms, with group members
indicated by the shaded area. In all cases, the group members form
distinctive peaks in velocity space.  Several groups show multiple
peaks, suggesting possible substructure. In particular, the two
richest groups, MZ\,10451 and MZ\,5383, both show two separate peaks.
However, the projected galaxy distribution shows no obvious spatial
segregation associated with these individual peaks to directly support
the existence of substructure. Recently, \citet{Hou09} concluded that
the Anderson-Darling (A--D) test is a reliable tool to detect
departures from a Gaussian velocity distribution in small data sets
with size typical of our groups. Because a non-Gaussian velocity
distribution could suggest an unrelaxed dynamical status, we performed
the A--D test on our groups and found that the null-hypothesis of a
Gaussian distribution is rejected in the two richest groups at the
90\% confidence level. This suggests that these two groups are not
relaxed systems. For the rest of the groups, the A--D test does not
reject the hypothesis of a Gaussian distribution.  However, the power
of the A--D test decreases in poorer systems \citep{Hou09}, and
consistency with a Gaussian velocity distribution need not imply a
dynamically relaxed systems in such cases.

\subsection{Morphology of the Groups}
\begin{figure}
\epsscale{0.8}
\plotone{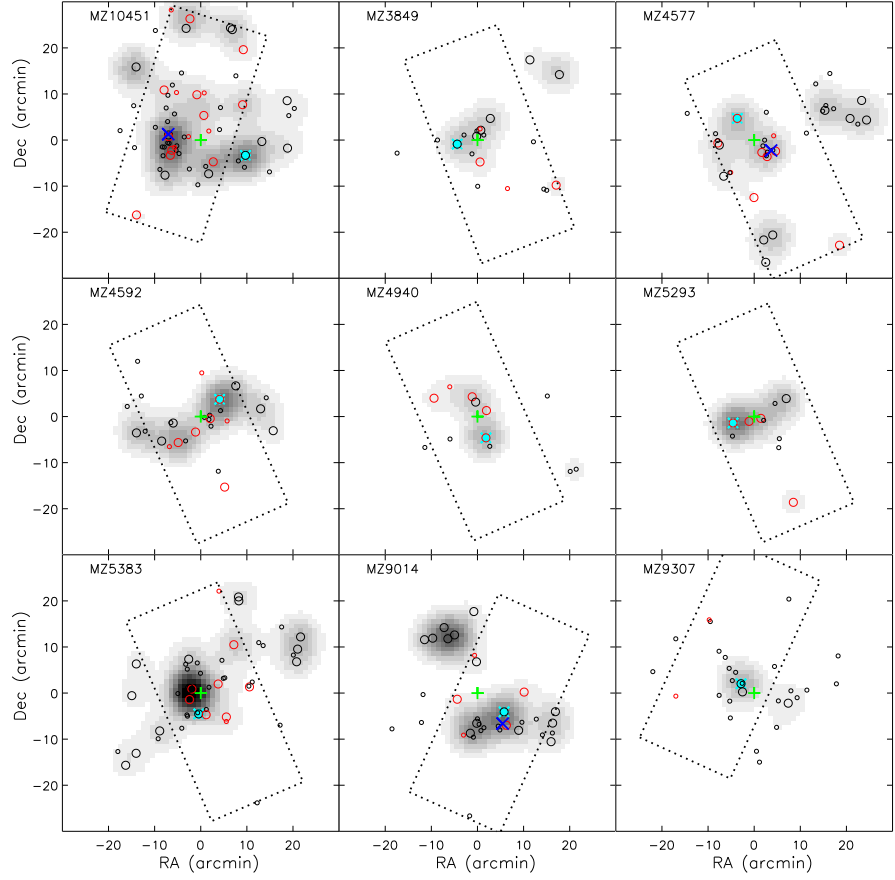}
\caption{Spatial distribution of the group galaxies. Big circles
  represent group members with $M_{R}\leq-20$ and small ones $M_{R}>-20$. Red
  circles are the sources with 24~\micron\ detection. Green plus signs
  indicate the luminosity-weighted group centers, and blue crosses
  denote the centroids of any detectable extended X-ray emission. The
  cyan stars are the brightest group galaxies within 1 Mpc of the
  group centers. Dotted rectangular regions show the MIPS 24~\micron\
  coverage in each field. Luminosity-weighted galaxy surface density
  maps are shown in grey, smoothed with a Gaussian kernel with FWHM =
  $7\arcmin$ ($\sim0.5$~Mpc).}
\label{f_galdist}
\end{figure}

In Figure~\ref{f_galdist} we show the projected spatial distribution
of the group members. Each plot is $1\deg\times1\deg$ in size, and the
dotted rectangular region indicates the MIPS 24~\micron\ coverage.
It is clear from the plot that the galaxy distribution in several
groups is rather irregular, with some of the groups showing
evidence of substructure. Rich systems such as MZ\,10451 and MZ\,5383,
tend to show clear concentrations in the galaxy
distribution, whereas several of the poorer ones display a more
filamentary overall morphology. However, the large uncertainty in
determining the geometry of the galaxy distribution in poor systems
renders such differentiation tentative at best. Furthermore, even the
richer systems show evidence of prominent subclumps and elongated
structures. We defer a detailed morphological analysis to a future
paper, when spectroscopic results for the full XI group sample are
available.

To determine the group center, we first calculated the $R$-band
luminosity-weighted center including all the group members.  We then
exclude galaxies with velocities more than 2$\sigma$ away from the
cluster mean or projected distance larger than 15 $\arcmin$ ($\sim$1
Mpc) from the initial group center and recalculate the
luminosity-weighted center.  We plot these centers as green plus signs
in Figure~\ref{f_galdist}.  We also overlay the luminosity-weighted
galaxy surface density maps in the same plot.  In many groups, as
shown in the plot, the luminosity-weighted group centers are quite far
away from the peak of the luminosity-weighted density map.  In the
three groups with clearly detected extended X-ray emission, MZ\,10451,
MZ\,4577 and MZ\,9014, the X-ray centroids are located closer to the
galaxy density peak ($<300$~kpc) than to the luminosity-weighted group
centers.
These large displacements may suggest that these systems are still in
the process of virialization.  Throughout this paper, we use
luminosity-weighted group centers as the group centers.  However, we
note that using luminosity-weighted galaxy density peaks (or X-ray
centroids, where available) as the group centers do not change the
general results in this paper.

Further evidence that many of these systems have not reached a fully
virialized stage is the lack of a central brightest group galaxy
(BGG). Nearby X-ray selected groups generally have early-type BGGs
sitting at the center of the group potential
\citep{Zabludoff98,Mulchaey98,Helsdon00}.  In most of our groups, the
brightest galaxies within 1 Mpc of the luminosity-weighted group
centers are located $>300$~kpc away from the centers.  Also, as shown
in Figure~\ref{f_galdist}, many of these BGGs are not coincident with
the most crowded region in the groups.  In the most X-ray luminous
group in the sample, MZ\,10451, the galaxy residing at the center of
the X-ray emission is only the fourth brightest galaxy in the
group. The BGG lies about 700~kpc away from the luminosity-weighted
center and is about 1.2~Mpc away from the center of the X-ray
emission, placing it beyond the virial radius according to the mass
profile determined from our X-ray data \citep{Rasmussen10}. In
addition, it has a radial velocity of $\sim 1000$~\kms\ relative to
the group mean and is a spiral galaxy.  These pieces of evidence
strongly argue that this BGG is just a recently accreted galaxy.
Moreover, the distribution of most of the bright galaxies in these
groups is fairly scattered instead of being centrally concentrated,
also suggestive of the systems generally not being dynamically old.

\section{Star-forming Properties of the Galaxies in XI Groups}\label{sec,results}
 
\subsection{Star-forming Galaxy Fractions}\label{sec,SF_frac}
A statistical indicator of star formation activity in a galaxy
population is the fraction of star-forming galaxies.  The sensitivity
of the 24 \micron\ observations of the XI groups allows the detection
of SF galaxies with SFR $>0.1$ \msunyr.  We can therefore calculate
the fraction of SF galaxies with SFR $>0.1$ \msunyr~for all the
galaxies brighter than $M_R=-20$.  We limit the SF fraction
calculation to galaxies with $M_R\leq -20$, because the comparison
cluster and field samples discussed in the following sections are only
complete down to this limit.  However, for the group galaxies, we
confirmed that extending the fraction calculation to a fainter
magnitude ($M_R\leq -19$) lowers the overall fractions but does not
change any of our conclusions. Dwarf galaxies fainter than $M_R=-19$
were excluded in the fraction calculations, due to incompleteness of
our optical spectroscopy (and, likely, of our 24\micron\ data) at
these magnitudes.
Since only group members within the region of 24~\micron\ coverage are
considered for fraction calculations, most galaxies at large radii are
also excluded.  We note that the following results remain unchanged if
we limit the fraction calculations to the galaxies covered by MIPS
within 1~Mpc from the optical group centers.

\subsubsection{Dependence of SF fractions on global group properties}
\begin{figure}
\epsscale{0.8}
\plotone{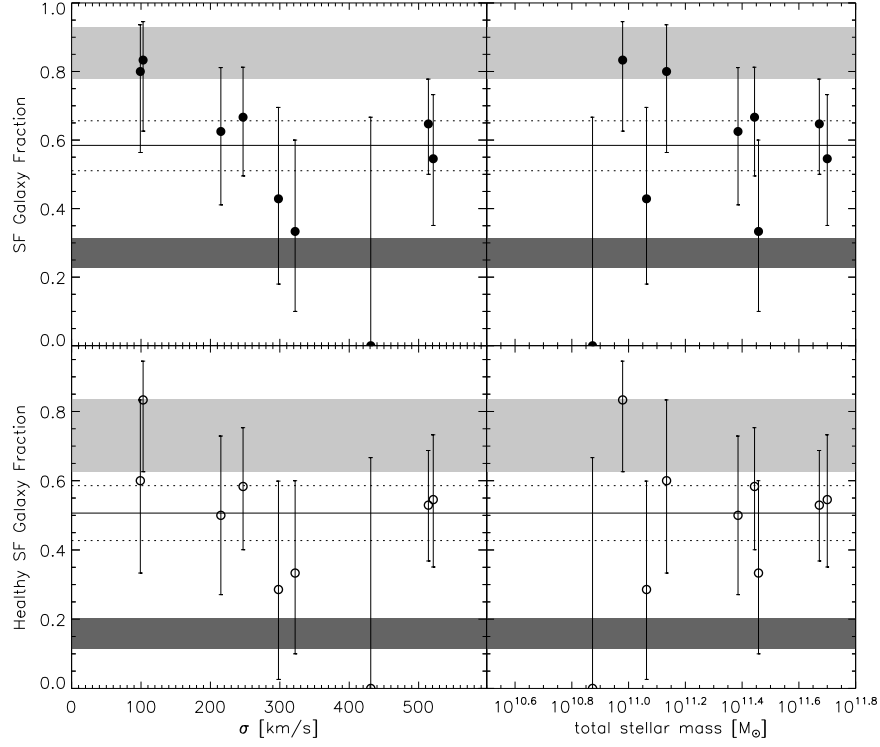}
\caption{Top: SF galaxy fraction vs.\ (left) group velocity dispersion
  and (right) total stellar mass of the groups. The fraction is
  defined as the ratio of SF galaxies with SFR\,$>0.1$~\msunyr\ to all
  galaxies with $M_R<-20$. Solid and dotted lines show the average
  fraction and its $\pm1{\sigma}$ range. Grey shaded regions indicate
  the $\pm1\sigma$ region of the average fractions of the field sample
  and dark shaded regions of the cluster sample.  Bottom: Same as the
  top panels but with the healthy SF galaxy fractions.  The healthy SF
  fractions is defined as SF galaxies with specific SFR more than 20\%
  of the typical specific SFR of the field galaxies from
  \citet{Salim07}.  }
\label{f_frac_depend}
\end{figure}

In the top left panel of Figure~\ref{f_frac_depend}, we plot the
star-forming galaxy fractions in the nine groups as a function of
group velocity dispersion.  Most of the groups (six out of nine) show
a high fraction ($>50\%$) of SF galaxies.  Two groups with
$\sigma\sim300$~\kms, MZ\,3849 and MZ\,9014, have smaller fractions of
43\% and 33\%, respectively. MZ\,9307 is a peculiar case with zero
fraction of star-forming galaxies: most of its members are dwarf
galaxies with only two galaxies brighter than $M_{\mathrm{R}}=-20$,
and these are not star-forming, resulting in a SF fraction of zero
with a large uncertainty.  A Spearman correlation test shows that
there is an anti-correlation between SF fractions and velocity
dispersion (at a significance of $93\%$), but this trend is mostly
driven by the two groups with smallest $\sigma$ ($\sim100$~\kms) and
highest SF fractions.  The significance of the anti-correlation drops
to $46\%$ once these two groups are excluded.

If the velocity dispersion of a group correlates well with the mass of
the group, which is true for a virialized system, then the lack of
clear trend in SF fractions with velocity dispersion suggests that the
SF properties of these groups do not depend strongly on total group
mass.  However, as evidenced by their irregular galaxy distribution
and the general lack of a central BGG, the groups in our sample are
not likely to be virialized, and their velocity dispersion might be a
poor indicator of their masses.  An alternative proxy for the group
mass is the integrated stellar mass of group members.  \citet{Yang05}
have demonstrated that the total stellar mass of group galaxies
brighter than $M_{\mathrm{R}}=-19.5+5\mathrm{log}h$ correlates tightly
with total group mass.  To derive the total stellar mass of the
groups, we used $R$-band magnitudes, calibrating the stellar mass
calculation using a subset of our group galaxies that have SDSS 5-band
photometric data.  For those galaxies, we can deduce their stellar
masses using the SED fitting methods proposed by \citet{Blanton07}.
The resulting stellar masses correlate well with the $R$-band
magnitudes, with a 1-$\sigma$ scatter of 0.17 due to variations in
stellar mass-to-light ratio.  We do not try to constrain the
mass-to-light ratio using galaxy colors, because group galaxies
outside our imaging region only have reliable $R$
magnitudes. Furthermore, the uncertainty on stellar mass caused by
such variations does not have a significant impact on the results in
this paper.

Using the inferred correlation, we deduced the stellar mass for group
galaxies in the full sample.  From the top right panel of
Figure~\ref{f_frac_depend}, it is clear that there is no correlation
between the SF fractions and the resulting total stellar mass of the
groups.  We also note that if we extend the SF fraction calculation to
include galaxies one magnitude fainter ($M_{\mathrm{R}}\leq -19$), the
SF fractions in all groups become smaller, but they still show no
trend with velocity dispersion nor with total stellar mass. As a
comparison, we also calculate the SF galaxy fraction for the field and
cluster samples.  In the field, this fraction is $87^{+6}_{-9}\%$,
about 30\% higher than the mean SF fraction of all group galaxies
($58^{+7}_{-7}\%$).  The two groups with the smallest velocity
dispersions ($\sigma\sim100$ \kms) have SF fractions consistent with
the field average.  The fraction in the cluster sample is
$27^{+4}_{-4}\%$, about 30\% lower than the group average.

We also note that the three groups with detectable extended X-ray
emission do not show different SF fractions compared to those with no
detection. Although the detection limits of our X-ray observations are
not uniform due to variations in the useful exposure time, the fact
that MZ\,10451, the most X-ray luminous group with a luminosity
typical of X-ray selected high-$\sigma$ groups, has an SF fraction of
$65\%$, very similar to the average of our nine groups, argues
against a significant correlation between SF fractions and the
X-ray properties of our groups.

The weak dependence of SF fractions on group global properties is
consistent with results from other studies of nearby groups
\citep{Balogh04} and groups at intermediate redshifts
\citep{Wilman05}.  This might be an indication that the SF properties
of group galaxies are more affected by their immediate environment,
e.g., local galaxy density, rather than the global environment.  It
could also suggest that the SF properties of the group members are not
directly related to their present environment \citep{Balogh04}.

\subsubsection{Stacked SF fractions}
\begin{figure}
\epsscale{0.8}
\plotone{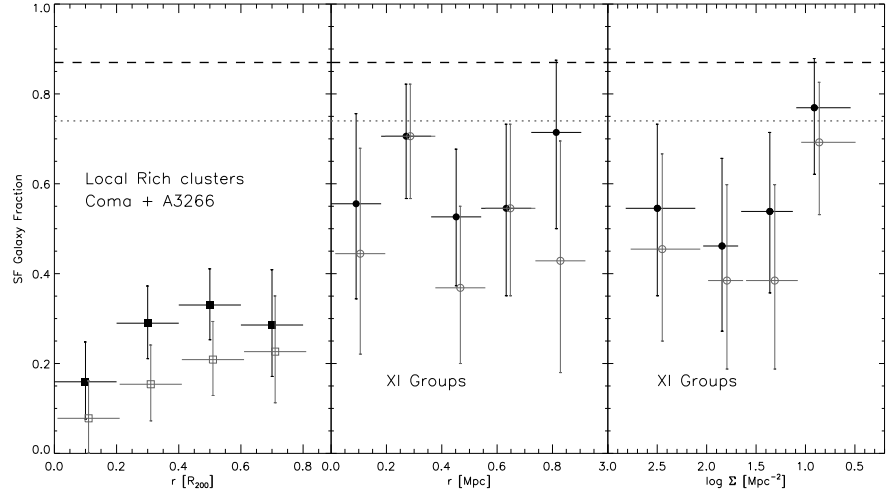}
\caption{Left panel: SF galaxy fractions (filled black squares) and
  healthy SF galaxy fractions (grey open squares) in the nearby rich
  clusters Coma and A3266 as a function of projected distance (in
  units of $R_{200}$) from the cluster center.  Middle panel: SF
  galaxy fractions (filled black circles) and healthy SF galaxy
  fractions (grey open circles) in groups as a function of projected
  distance from group centers.  Right panel: SF fractions
  and healthy SF galaxy fractions in groups as a function of local projected
  galaxy density.  In all three panels, the black dashed line is the
  average SF fractions in the field sample and the grey dotted line is
  the average healthy SF fractions in the field sample.  }
\label{f_frac}
\end{figure}

In nearby rich clusters, \citet{Bai09} showed that the local SF
fraction increases linearly from the cluster core to large radii (see
also \citealt{Mahajan10}).  For individual groups in our sample, there
are too few galaxies to study this trend, so we stack all the groups
together and plot the SF fraction against the projected distance from
group luminosity-weighted centers in the middle panel of
Figure~\ref{f_frac}. We note that not all annuli in this plot are
uniformly covered by MIPS in all groups, but since only group members
within the region of MIPS coverage are included in the fractions,
partial coverage should not introduce systematic variations in the
plot. In contrast to the rich clusters (shown in the left panel of
Figure~\ref{f_frac}), SF fractions in groups show no clear dependence
on the distance from the group centers and remain at a level higher
than the outer region of the rich clusters ($> 0.5R_{200}$, $R_{200}$ is the radius within which the mean cluster density is 200 times the critical density of the universe at that redshift).  Using a
larger bin size does not change this result.  We note that the left
panel of Figure~\ref{f_frac} is similar to figure~7 of \citet{Bai09},
but with SFRs updated according to the prescription in
Section~\ref{sec:SFR}.  For our groups, some of which are likely not
virialized, the virial radius is non-trivial to evaluate and may not
be a meaningful measure, so we do not normalize the radii by $R_{200}$
before stacking the group results.

The continuously decreasing SF fractions toward the cluster center
could reflect a dependence of the SF properties on cluster properties
that themselves depend on radius, such as local galaxy density or the
density of the intracluster medium (ICM). If this is the case, the
lack of a dependence of SF fractions on projected radius in groups
could be a result of a breakdown of the correlation between galaxy
density and projected distance rather than a breakdown of the
correlation between SF fractions and galaxy density.  The apparent
displacements between the luminosity-weighted group centers and the
galaxy density peaks seem to support this argument (see
Figure~\ref{f_galdist}).  To check this possibility, we calculated the
projected local galaxy density ($\Sigma$) for all the group galaxies
using the distance to the 2nd nearest neighbor ($r_{2d}$),
$\Sigma=3/(\pi r_{2d}^{2})$.  After assigning a local density to each
group galaxy, we calculate the SF fractions in four density bins for
all the galaxies brighter than $M_{R}=-20$ and within $10\arcmin$
(0.7~Mpc) away from the group centers.  The spectroscopic survey is
less complete in the outer regions ($>10\arcmin$) which could give
rise to a systematically lower local density for galaxies at those
radii.  The density bins are selected to have roughly the same number
of galaxies in each bin, and the highest density bin has density
comparable to the value at 0.2$R_{200}$ of the clusters.  In the right
panel of Figure~\ref{f_frac}, we plot the SF fractions as a function
of density.  However, the SF fractions of all three high density bins
remain approximately constant at a level higher than the fractions in
the cluster.  The lowest-density bin has a marginally significant
higher SF fraction, and it is consistent with the average field value.
This could be evidence of an anti-correlation between density and SF
fractions, but the large uncertainties preclude robust conclusions in
this regard (51\% probability from a Spearman test).

\subsection{Specific SFR of Group Galaxies and Healthy SF Galaxy Fractions}{\label{s_ssfr}}
In the previous section, we focused on the study of the total current
SFR in a galaxy.  However, a fixed SFR can contribute substantially to
the total stellar mass, color, and optical spectrum of a small galaxy
but little to a massive one. In this sense, the specific SFR, i.e.\
the total SFR normalized by the stellar mass of a galaxy, is a better
measure of the relative importance of SF in different galaxies.
Because the stellar mass of a galaxy is the integral of the past SFR,
the specific SFR, to first order, is also a measure of star formation
history.  Many studies \citep{Gavazzi96, Boselli01, Kauffmann04} have
indicated that the specific SFR depends primarily on galaxy mass, with
any environmental dependence being of secondary importance.  To
disentangle the environmental dependence from the mass dependence, we
derived the specific SFRs of our group galaxies and compared them to
those of cluster and field galaxies.

\begin{figure}
\epsscale{0.8}
\plotone{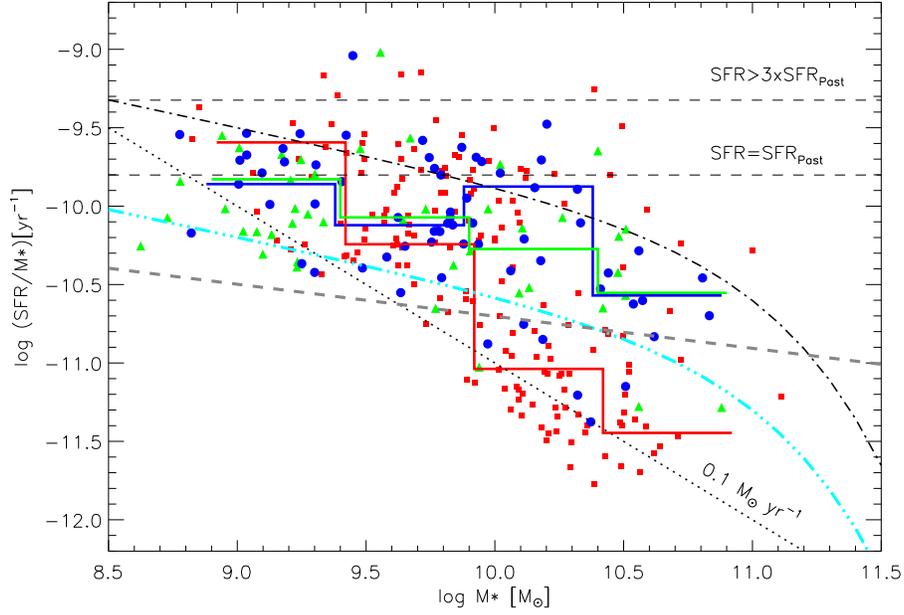}
\caption{Specific SFRs (SFR/$M_\ast$) of galaxies as a function of
  stellar mass. Filled blue circles, green triangles, and red squares
  represent group, field, and cluster galaxies, respectively.
  Black dash-dotted curve indicates the most frequent specific SFRs of
  field galaxies as determined by \citet{Salim07}, and the cyan
  dot-dot-dot-dash line represents 20\% of this value, used to define
  ``healthy'' SF galaxies.  Blue, green, and red histograms show the
  modes of the specific SFRs for SF galaxies with SFR\,$>0.1$~\msunyr\
  in our group, field, and cluster sample, respectively.  The black
  dotted line shows the $0.1$~\msunyr\ detection limit of our
  24~\micron\ group data.  The grey dashed line is the SF dividing
  line for passive/active galaxies in \citet{Weinmann06}.  The two
  black dashed horizontal lines correspond to birthrates of 3 and 1.
  }
\label{f_ssfr}
\end{figure}

In Figure~\ref{f_ssfr}, we plot the specific SFRs of group galaxies as
a function of their stellar mass.  The star-forming group galaxies
with $M_\ast<10^{10.5}M_{\sun}$ concentrate, although with a fair
amount of scatter, on a star-forming sequence
\citep[e.g.,][]{Salim07}. The more massive galaxies generally show
lower specific SFRs.
Along with group galaxies, we plot the star forming sequence of a
large sample of local SF galaxies from \citet{Salim07}.  Using
ultraviolet (UV) data from {\em GALEX} along with SDSS data,
\citet{Salim07} measured the SFRs of $\sim$\,50,000 nearby galaxies
and calculated the mode, i.e.\ the value that occurs most often, of
the specific SFR as a function of stellar mass. They characterized
this function with a Schechter function, plotted as a dash-dotted
curve in Figure~\ref{f_ssfr}.  It shows for each stellar mass the most
frequently observed specific SFRs of field galaxies. Although
\citet{Salim07} used a different SFR indicator, their UV-inferred and
IR-inferred SFRs show good overall agreement, with a systematic offset
of just $\sim0.02$~dex and scatter of $\sim0.5$~dex \citep{Salim09}.
The difference between their IR SFR estimation and ours introduces a
systematic difference $<0.1$~dex in the SFR range of our interest
\citep{Salim09}.  To compare with their SF sequence, we also calculate
the mode of the specific SFRs for the group galaxies, as well as the
modes for our field and cluster samples, in four stellar mass bins.
To be consistent, we only calculate the mode for SF galaxies with
SFR\,$>0.1$ \msunyr, which is the detection limit of the group
galaxies.  This detection limit is higher than that of the UV data
used by \citet{Salim07}, which are sensitive down to at least
SFR\,$\sim0.01$ \msunyr.  Therefore, the modes of the specific SFRs of
the galaxies calculated here only represent the typical value of the
upper envelope in the distribution of specific SFR vs.\ stellar mass,
and they could overestimate the actual typical specific SFRs.

The modes of our field sample, although subject to large uncertainties
due to the small sample size, generally follow the SF sequence found
by \citet{Salim07}, but are systematically lower by 0.3~dex.  Part of
this difference comes from the systematic difference in SFR estimates
($\sim0.1$~dex).  This systematic difference is not significant
though, given the intrinsic scatter in specific SFR along the
star-forming sequence \citep[0.5 dex,][]{Salim07} and the scatter in
the correlation between the UV- and IR-inferred SFRs.  At
$M_\ast<10^{10}\msun$, the SF galaxies in clusters, groups, and the
field occupy similar regions in the plot and do not show any
significant difference in their distribution.  However, in this mass
range, the 24 \micron\ sensitivity limits us from detecting SF
galaxies with relatively low specific SFRs. The similarity therefore
only shows that the upper envelope of the specific SFR vs.\ stellar
mass distribution is broadly similar in all environments within this
mass range.  However, it is not clear whether there is a different
distribution for SF galaxies below our detection limit.  At
$M_\ast>10^{10}\msun$, the typical specific SFRs of the XI group
galaxies are still very similar to those of the field.  However, the
cluster SF galaxies show much lower typical specific SFRs compared to
group and field galaxies of similar mass.  This difference is mostly
due to a fraction (14\%) of massive galaxies with very low specific SFR
$<10^{-11}$~yr$^{-1}$ that is the most prominent in clusters. 
The SFRs of these
galaxies are very close to our detection limit, and their 24 \micron\
luminosities are only a few times higher than what is expected from
the cold dust emission of an old stellar population.
The SF properties of these massive galaxies are very similar to 'dusty red galaxies' found in the A901/2 super cluster \citep{Wolf09,Gallazzi09}. In order to find out if the massive SF galaxies are the same population as the dusty red galaxies, we compare their colors and morphologies with the galaxies in the Coma cluster where we have this information. 
We confirm that the majority of the massive SF galaxies are indeed red in color.
However, the morphologies of these galaxies are typically S0/Es, differing from the dust red galaxies discussed by \citet{Wolf09}, which are mostly passive spiral galaxies (Sa/Sb).

\begin{figure}
\epsscale{0.8}
\plotone{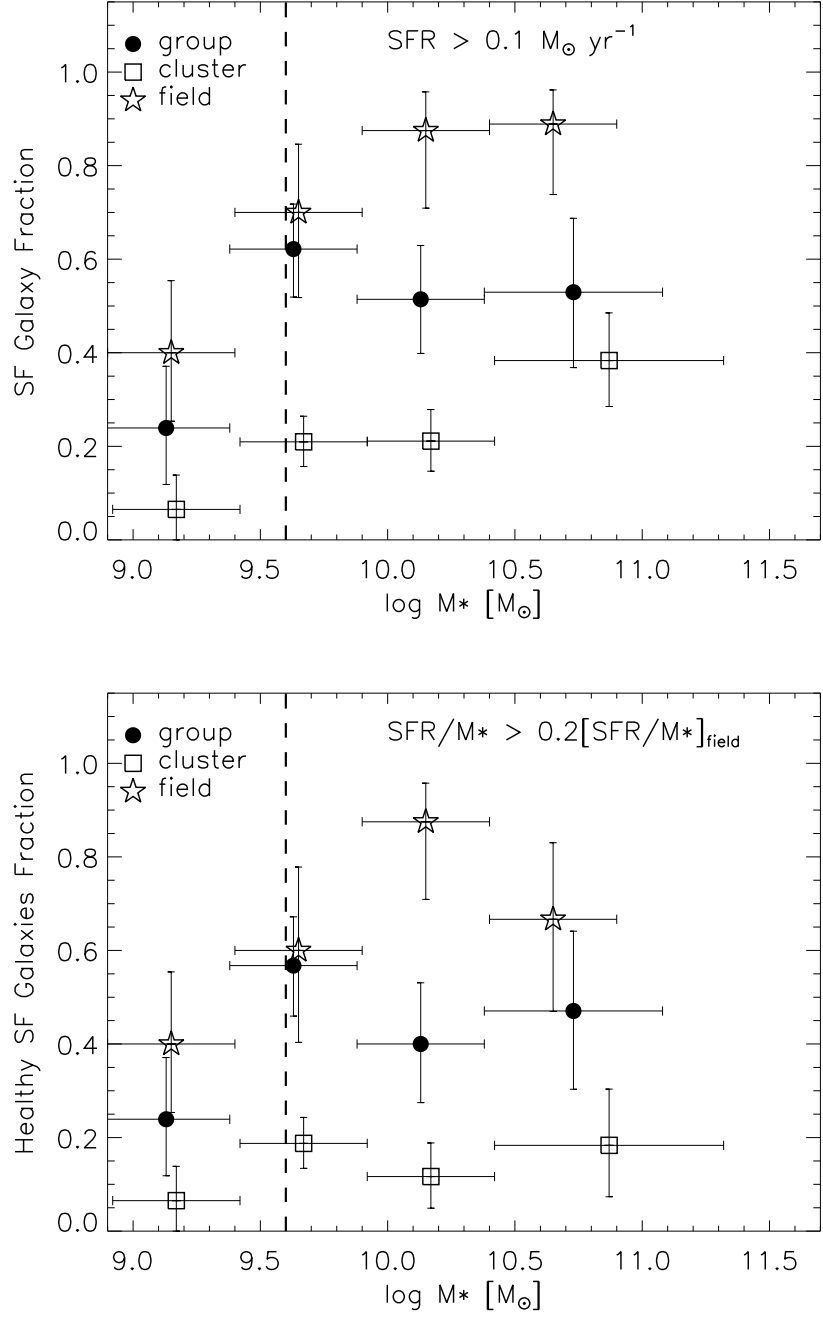}
\caption{Top panel: SF galaxy fraction as a function of stellar
 mass. Filled circles, open stars, and open squares show the group,
 field, and cluster galaxies, respectively.  Bottom panel: As above,
 but for the healthy SF galaxy fractions.  Dashed vertical lines are
 shown at $M_\ast = 10^{9.6}$ \msun, corresponding to $M_{R}=-20$.  }
\label{f_frac_mass}
\end{figure}

In the top panel of Figure~\ref{f_frac_mass}, we show the fractions of
galaxies with SFR\,$>0.1$ \msunyr\ in four stellar mass bins.  For
both groups and the field, the fractions of galaxies with SFR\,$>0.1$
\msunyr\ are all $>50$\% at $M_\ast>10^{9.5}\msun$.  This high
detection fraction helps to support the robustness of the typical
specific SFRs derived for these two samples in this mass range.  In
the lowest mass bin, the detection fractions in all environments are
quite low, and we begin to lose SF galaxies on the SF sequence due to
the detection limit.  In this mass bin, the derived specific SFR modes
only represent the typical value of a subset of the SF galaxies.  In
all mass bins, the SF fractions in groups are lower than those in the
field, with the clusters showing the lowest fractions.  For galaxies
with $M_\ast>10^{10.5}$\msun, the SF fraction in clusters is
comparable to that in the groups.  However, it is clear from
Figure~\ref{f_ssfr} that the typical specific SFR of cluster galaxies
in this mass range is lower than the value of the group and field
galaxies; definining SF galaxies using a fixed SFR limit for galaxies
of differing mass clearly fails to distinguish these different
populations of SF galaxies.  To take this mass dependence of the
specific SFRs into account, we assume that the typical specific SFR of
field galaxies is the value for a galaxy unaffected by its
environment, and define \lq \lq healthy\rq \rq \ SF galaxies as
galaxies with specific SFR at least 20\% of the typical value of field
SF galaxies of the same mass, as found by \citet{Salim07}.
If we take into consideration the systematic difference between the SF
sequences obtained from our field sample and from the SDSS SF galaxy
sample with UV data, and assume that the average scatter in the SF
sequence is the same as that of the SDSS SF galaxies of
\citet{Salim07}, then 75\% of the SF galaxies in our field sample
should have specific SFRs above this limit.  In the bottom panel of
Figure~\ref{f_frac_mass}, we show the healthy SF galaxy fractions as a
function of stellar mass.  In the three high-mass bins, where we are
sensitive to all the healthy SF galaxies, the fractions in our field
sample are all above 60\% and generally consistent with the 75\% from
the above expectation.  The fractions in groups range from 40-60\%,
all lower than in the field.  But the difference is most pronounced in
the $M_\ast = 10^{9.5-10.4}$\msun\ bin and is not significant in the
other mass bins.  In clusters, the fractions remain very low in all
mass bins, 10-20\%.  Although the SF fraction in clusters is about
40\% at $M_\ast >10^{10.5}\msun$, almost as high as that in groups,
the healthy SF fraction is only 20\%, much lower than that in groups.

It is clear from Figure~\ref{f_ssfr} and Figure~\ref{f_frac_mass} that
using a fixed SFR cut to define SF galaxies and assess the SF
properties in different environments may not be able to differentiate
massive galaxies with lower than typical SFR from low-mass normal SF
galaxies.  Therefore, we repeat the analysis of
Section~\ref{sec,SF_frac} by calculating the healthy SF galaxy
fraction for each group.  We limit the calculation to galaxies with
$M_\ast>10^{9.6}\msun$, which corresponds to the $M_{R}=-20$ magnitude
cut used for calculating the SF fractions.  Above this limit, our data
are sensitive to all the healthy SF galaxies.  In the bottom panels of
Figure~\ref{f_frac_depend}, we plot the healthy SF galaxy fractions
against group velocity dispersion and total stellar mass.  Again,
there is no obvious systematic trend between healthy SF fraction and
velocity dispersion or total stellar mass.  The probability of an
anti-correlation between healthy SF fractions and velocity dispersion
is only 84\%.  The average fraction in groups is $51^{+8}_{-8}\%$,
lower than that of the field ($74^{+9}_{-11}\%$) but higher than in
clusters ($16^{+4}_{-4}\%$).

\begin{figure}
\epsscale{0.6}
\plotone{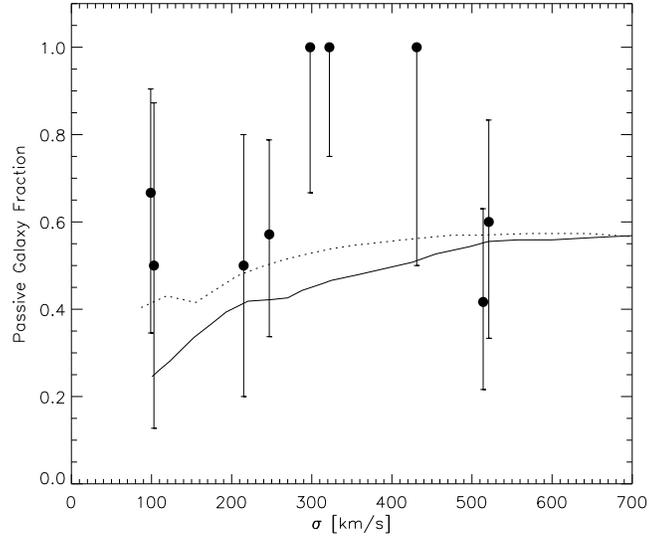}
\caption{Passive galaxy fractions as a function of group velocity
 dispersion. The passive galaxies are defined by the criteria in
 \citet{Weinmann06} and the curves are the trends found by these
 authors using an SDSS group catalog.  The dotted curve is derived by
 estimating SDSS group masses from velocity dispersion, and the solid
 line from total luminosity \citep{Weinmann06}. }
\label{f_frac_et}
\end{figure}

The lack of dependence of healthy SF fractions on group velocity
dispersion and total stellar mass would appear to be in contrast with
the study of \citet{Weinmann06}.  Using a galaxy group catalog
constructed from SDSS data, \citet{Weinmann06} divide galaxies into
early-type and late-type galaxies, based on criteria very similar to
what we use to define healthy SF galaxies, but they found that the
fraction of early-type galaxies in groups increases with group halo
mass.  Specifically, they define early-type galaxies as both red in
color and ``passive'' in terms of their star forming properties.
Because only 1\% of their group population is blue and ``passive'',
the early-type galaxies are approximately equivalent to passive
galaxies.  They adopt a mass-dependent specific SFR cut to divide
passive and active galaxies, which is very similar to our criteria for
healthy SF galaxies (shown as the grey dashed line in
Figure~\ref{f_ssfr}).  To directly compare with their results, we use
the same SFR criteria to define passive galaxies in our group sample
and calculate the passive galaxy fractions for galaxies with $M_\ast
>10^{10}\msun$, equivalent to the magnitude cut of \citet{Weinmann06}.
In Figure~\ref{f_frac_et}, we show this fraction as a function of
group velocity dispersion.  Again, similar to what we find in terms of
healthy SF fractions, the fractions of passive galaxies in our group
sample do not show any strong trend with velocity dispersion.  A
Spearman test shows that the probability of correlation is only 25\%.
This apparent discrepancy could be due to small number statistics
owing to the limited size of our group sample. In fact, within the
rather large statistical uncertainties, the majority of our nine
groups do show fractions consistent with the mean trends found by
\citet{Weinmann06}.  In addition, the discrepancy could arise partly
from differences in the adopted group definitions.  \citet{Weinmann06}
select groups using a halo-based group finder \citep{Yang05} while the
MZ sample is based on a friends-of-friends algorithm.  The halo-based
group finder assumes that groups are virialized systems, but this
assumption is unlikely to hold in general for the groups in our
sample, where some systems are probably still in the process of
collapsing and so are not yet fully virialized.

Finally, we also examined the healthy SF fractions as a function of
radius from the group centers and the local projected galaxy density.
These fractions are shown as grey squares in Figure~\ref{f_frac}.
Although the healthy SF fractions in all environments are lower than
the ``standard'' SF fractions, the general results obtained earlier
remain unchanged: there is no systematic dependence of healthy SF
fractions on the distance from the group centers. although there might
be a weak anti-correlation between healthy SF fractions and projected
galaxy density.  The healthy SF fractions of groups are higher in all
density bins than the fractions seen in the outskirts of clusters.

\subsection{SF History of Group Galaxies}
As mentioned, the current specific SFR of a galaxy provides a rough
measure of its star formation history.  The specific SFR can be
directly related to the birthrate, which is defined as the current SFR
normalized by its past average.  Given a gas recycling fraction $R$
and stellar age $\tau$, the birthrate $b=\frac{SFR}{\langle
SFR_{past}\rangle}=\frac{SFR\cdot\tau}{M_\ast}(1-R)$
\citep{Kennicutt94}.  Assuming $R=0.5$ and the age of the universe at
$z=0.06$ as an upper limit to $\tau$ \citep{Brinchmann04}, we can
calculate the upper limit of the birthrate for group galaxies.  As
shown in Figure~\ref{f_ssfr}, most of the SF group galaxies with
$M_\ast<10^{10.4}M_{\sun}$ have $b\approx1$, indicating they are
forming stars more or less at the same rate as the past average.  If
we define galaxies with current SFRs three times larger than the past
average as starburst galaxies \citep{Brinchmann04}, only one group galaxy fulfills this
criterion.  For more massive group galaxies, their current SFRs are
generally lower than their past average, consistent with the trend for
field galaxies.  In clusters, there is a relatively large number of SF
galaxies with current SFR significantly lower than their past average.

\section{DISCUSSION}\label{sec,discuss}
\subsection{The SF Fraction in Different Environments}
On average, the SF fraction in our group sample is 30\% lower than
that in the field.  Even though this deficit of SF galaxies may be
minimal in the least massive systems and the lowest density regions,
it prevails in the typical environment of our group population.  This
result is consistent with the higher fraction of passive galaxies
found in many different group samples at different redshift ranges
compared to the field \citep{Balogh04, Wilman05,Jeltema07,Balogh09}.
On the other hand, when compared to the clusters A3266 and Coma, the
SF fraction in groups is generally $30\%$ higher.  The decreasing SF
fractions in more massive structures could suggest an environmental
effect, but it does not directly imply an environmental suppression of
the galaxy SF.  Many studies have shown that massive galaxies formed
their stars early on in a short time period while less massive
galaxies evolved more gradually \citep[e.g.,][]{Cowie96,Heavens04}.
This suggests that the stellar mass is probably the most important
variable that regulates the star formation history of a galaxy
\citep[e.g.,][]{Noeske07}. If galaxy clusters and groups, which
originated from the highest density perturbations in a cold dark
matter (CDM) universe, also preferentially host more massive galaxies
\citep{Bardeen86}, then the observed difference in the SF properties
between high- and low-density environments was seeded at the beginning
of the Universe.  This is considered to be the ``nature'' scenario of
galaxy evolution.  However, the fact that the healthy SF galaxy
fractions, which by definition already take into account the stellar
mass dependence of the SF properties, also show an increasing deficit
in denser environments which persists across a large stellar mass
range, suggests this scenario cannot fully explain those differences
and that some additional ``nurture'' process that invokes
environmental effects is needed \citep{Christlein05,Baldry06}.

\subsection{Comparing Groups with the Field}
Even though the group galaxies have, on average, a much lower SF
fraction than the field, the typical specific SFRs of group SF
galaxies are not very different from those of field galaxies at 
$M_\ast>10^{9.6}\msun$.  Similar to our result, \citet{Tyler10} also found 
the specifc SFRs of group and field galaxies of intermediate redshift ($0.3<z<0.5$) are not very different. 
Therefore, if the much lower SF fractions in 
groups is caused by some environmental mechanism, the transformation
needs to happen on a short time scale.  Such a mechanism should
quickly turn normal SF galaxies into quiescent ones without affecting
the overall typical specific SFRs too much.  Galaxy-galaxy
interactions appear to be a likely candidate for this mechanism.
These interactions can trigger starbursts in galaxies and cause them
to exhaust their gas fuel quickly \citep{Mihos94}.  This mechanism is
expected to be most efficient in poor groups.  The tentative
dependence of SF fractions on the local galaxy density also
favors galaxy-galaxy interactions as the dominant mechanism.

\citet{Miles04} found a prominent dip at $M_{R}\approx-19.5$ in the
optical luminosity functions of poor groups and attributed it to
galaxy merging by dynamical friction, which preferentially depletes
the intermediate-luminosity galaxies. 
Corresponding to the dip, there is also a bump at brighter magnitudes
($M_{R}\approx-20.5$), which can be explained by the boosted galaxy
number resulting from merged lower-mass galaxies.  Interestingly, the
healthy SF fraction in our group sample also shows a potential dip at
$M_\ast\approx 10^{10.1}\msun$.  In this stellar mass bin, the group
sample shows the largest deviation from the field sample, with
$\sim40\%$ fewer healthy SF galaxies.  In the two neighboring mass
bins, the fractions in the group sample are still lower than that in
the field, but the differences are not significant. Incidentally, the
dip of the healthy SF fraction occurs roughly in the stellar mass
range where the bump of the optical luminosity function is located
($M_\ast\approx10^{9.8}$\msun).  This supports a scenario in which
there are relatively more merged galaxies in this stellar mass range
which have exhausted their gas fuel at the early starburst stage of
the interaction and now show little SF.  However, a larger sample of
group and field galaxies would be needed to confirm this possible dip
in healthy SF fractions.

If such interaction--triggered starbursts are responsible for eventual
SF suppression, the detection of galaxies with enhanced star formation
could be direct evidence of such processes in action.  However, we
only detect one such galaxy in our sample.  Similar to our work,
\citet{Balogh09} also failed to detect galaxies with enhanced star
formation in their intermediate redshift group sample.  However, the
detectability of such a population in groups depends sensitively on
the duration of the interaction-induced starburst.  Using
hydrodynamical simulations, \citet{Cox06,Cox08} investigated
properties of merger-driven starbursts and the effects of employing
different supernova feedback models.  Although their simulations
suggest that the amount of the total gas consumption of the
merger-induced star formation that is directly related to SF
suppression is invariant with respect to the choice of feedback model,
the duration of the starbursts does vary from model to model.  Their
``stiff'' feedback model assumes the star-forming gas has an equation
of state $P\sim \rho^{2}$ while their ``soft'' model assumes the gas
is isothermal.  The stiff feedback model predicts a starburst
timescale of $\ge0.5$ Gyr, whereas the soft model results in a much
short timescale $\ge0.1$ Gyr.  If we adopt the time scale given by the
stiff feedback model, the starburst fraction of $<1$\% found for the
present group sample sets an upper limit of 10\% for the fraction of
galaxies that have experienced such a merger-induced starburst within
the last 5~Gyr.  This fraction falls short of explaining the 30\%
deficit of SF galaxy found in current groups population compared to
the field.  However, if we adopt the shorter timescale given by the
soft feedback model, enough merger-induced starbursts will have
occurred within the last 2--4~Gyr to explain the deficit.

In addition to galaxy-galaxy interactions, other mechanisms could also
be at work within the group environment.  Even though the gas density
of the intragroup medium is usually not sufficient to ram pressure
strip the cold disk gas in a galaxy, it could efficiently strip any
hot gaseous halo \citep{Rasmussen06a,Kawata08,McCarthy08}.  This would
cut off the supply that replenishes cold gas and eventually shut down
the star formation in galaxies. The time scale of this quenching
process, so-called strangulation, is much longer ($>$~1~Gyr).  If it
is the dominant process responsible for the deficit of SF galaxies, we
would expect to see many SF galaxies with suppressed SFRs.  This is
inconsistent with the similar typical specific SFR we found in our
group and field samples across a large stellar mass range, suggesting
that strangulation is probably not the dominant mechanism for galaxies
with $M_\ast>10^{9.6}\msun$ in the poor groups studied here.  In X-ray
luminous groups, however, this mechanism may become more important.

\subsection{Comparing Groups with Clusters}
In hierarchical structure formation, galaxy clusters are assembled
from lower mass halos and the galaxies in clusters might have been
residing in group environments before they finally fell into clusters.
This makes ``preprocessing'' in groups potentially important for
cluster galaxies \citep{Zabludoff98}.  However, the importance of
preprocessing depends on the accretion history of clusters.  Using CDM
$N$-body simulations, \citet{Berrier09} claimed that the majority of
cluster galaxies (70\%) have never resided in a group environment
before they fell into the cluster, and therefore that preprocessing in
group environments could not play a significant role in explaining the
difference between cluster and field galaxies.  This result seems to
be at odds with the observation that about half of the galaxy
population resides in group environments in the nearby Universe
\citep{Eke04,Eke05,McGee09}, because it would mean that clusters
preferentially accrete isolated galaxies rather than galaxies in
groups.  The apparent discrepancy, again, is likely related to
differences in the adopted group definition.  \citet{Berrier09} define
group members as the halos within the virial overdensity boundary of
the hosting dark matter halos.  This definition results in a much
smaller fraction ($\sim15\%$) of galaxies residing in group
environments compared to those found by the friends-of-friends
algorithms typically used to identify groups in observational samples.
Given the small fraction of galaxies residing in groups according to
the former definition, it is no surprise that the majority of the
galaxies falling into clusters have never been preprocessed within a
group environment by this definition.  On the other hand, if we follow
the much more relaxed group definition as adopted in
friends-of-friends algorithms, we would expect a much higher fraction
of cluster galaxies to have been part of the group environment prior
to falling into clusters.  However, even if all the cluster galaxies
have been in groups, the lower fraction of SF galaxies in clusters in
a large stellar mass range compared to the fraction found in the XI
groups suggests that preprocessing in groups is not sufficient to
explain the deficiency of SF galaxies in clusters.  This is reinforced
by the fact that even in the highest density regions of our groups,
the SF fraction is still higher than the fraction found in the
outskirts of clusters.  To explain the low SF fractions in clusters,
either an environmental mechanism that works in group environments
must continue to work in clusters, or some other cluster-specific
environmental mechanism must be invoked.  In either case, further
processing of SF galaxies within the cluster environment is required.

Not only are the SF fractions of cluster galaxies on average smaller
than in groups, but the SF galaxies in clusters have smaller specific
SFRs at $M_\ast>10^{10}\msun$.  Such differences strongly support
further SF suppression in clusters.  \citet{Bai06,Bai09} found that
the IR luminosity function of nearby rich clusters has the same shape
at the bright end ($L_{\rm IR}>10^{43}$~\ergss) as that of field
galaxies.  The galaxies that contribute to the bright end of the IR
luminosity function are galaxies with SFR\,$>0.3 \msunyr$, which
constitute the upper envelope of the specific SFR vs stellar mass
distribution of cluster galaxies (cf.\ Figure~\ref{f_ssfr}).  The
similarity of the bright-end shape of the IR luminosity function in
different environments is corroborated by the similar upper envelope
of the specific SFR vs stellar mass distribution found here.
\citet{Bai09} suggest this similarity points to a fast acting SF
suppression mechanism in clusters, for example, ram pressure
stripping, that produces few galaxies in transition.  However, the
comparison of the specific SFRs of cluster SF galaxies to those in the
groups and field does reveal a population of massive SF galaxies
($M_\ast>10^{10}\msun$), with suppressed but not totally extinguished
SF, predominantly seen in clusters. 
This difference does not necessarily exclude ram pressuring stripping as an important mechanism
in clusters, however, because massive galaxies are less vulnerable to
such processes and could still retain some of their gas following a
stripping event.  The residual gas in those massive galaxies could
sustain low-level SF for a much longer time.

\section{SUMMARY}\label{sec,summary}
We have presented the first mid-IR study of nearby groups with
complete optical spectroscopy and X-ray data. The nine groups in our
sample span a wide range in velocity dispersion (100--500~\kms), X-ray
properties, richness, and galaxy distribution.  These groups are
typical of the galaxy groups that make up more than half of the galaxy
population in the nearby Universe, and they are likely covering a wide
range of evolutionary states. We analyzed the SF properties of the
group galaxies from their MIPS 24~\micron\ emission and tested for
correlations with global group properties. We also compared the SF
properties of the group galaxies with those of cluster and field
galaxies at the same redshifts to investigate the environmental effect
on galaxy evolution. Our major results are summarized as follows:

1) The projected galaxy distributions of the nine groups show large
 variations, from more concentrated and circular distributions in the
 most massive groups ($\sigma\sim500$~\kms) to filamentary structures
 in the least massive ones ($\sigma\sim100$~\kms).  This variation,
 along with the lack of a dominant BGG in the group center and the
 generally low level of extended X-ray emission, suggests that some of
 these systems are not yet virialized but still in the process of
 collapsing.

2) On average, the SF galaxy fraction (SFR\,$>0.1\msunyr$,
 $M_{R}<-20$) in our group sample is about 30\% lower than in a
 comparison field sample extracted from the same data, and 30\% higher
 than in our comparison cluster sample.  The SF fraction of our groups
 does not show a strong systematic dependence on group global
 properties such as velocity dispersion, total stellar mass, or the
 presence of detectable diffuse X-ray emission.
 However, the two groups with the smallest velocity dispersion
 ($\sigma\approx100$~\kms) do show the highest SF fractions, at a
 level comparable to that of the field.  These conclusions remain
 unchanged if only considering the ``healthy'' SF galaxy fractions
 (SFR/$M_\ast>$0.2[SFR/$M_\ast$]$_{\rm field}$,
 $M_\ast>10^{9.6}\msun$).

3) There is no strong dependence of the SF fraction on the radial
 distance from the group center.  The 24~\micron\ SF fraction in the
 groups is, at all radii, larger than the corresponding fraction in
 the outskirts ($\sim R_{vir}$) of rich clusters at similar redshifts.
 There is a weak trend of SF fractions decreasing with increasing
 projected galaxy density, with the lowest density regions having an
 SF fraction comparable to the field population.  Even in the highest
 density regions of groups, the SF fraction is still larger than the
 SF fraction in the outer regions of clusters.  In addition, the
 healthy SF fractions of cluster galaxies across a large stellar mass
 range are all at least 20\% lower than those in groups.  These pieces
 of evidence strongly suggest that preprocessing of galaxies in group
 environments prior to infall into clusters is not a sufficient
 explanation for the lower fraction of SF galaxies in clusters and
 that further processing by the cluster environment is required.

4) The typical specific SFRs of SF galaxies in groups are very similar
 to that in the field across a wide mass range
 ($M_\ast>10^{9.6}\msun$), favoring a quickly acting mechanism that
 suppresses star formation to explain the overall smaller fraction of
 SF galaxies in groups.  The healthy SF fractions in groups show a
 possible dip at $M_\ast \approx 10^{10.1}\msun$, corresponding to the
 bump seen in the optical luminosity function of poor groups
 \citep{Miles04}.  This is consistent with the speculation that galaxy
 merging by dynamical friction preferentially depletes
 intermediate-luminosity galaxies, which become subject to rapid gas
 consumption during the interaction, resulting in a population of
 galaxies of high stellar mass and a relatively low fraction of
 healthy SF galaxies. If galaxy-galaxy interactions are responsible
 for the deficit of SF galaxies in groups, then our non-detection of a
 significant starburst population among current group members does
 indeed imply a short time-scale for any merger-induced starburst
 stage ($\sim0.1$ Gyr). This agrees well with a supernova feedback
 model that assumes an isothermal state for the star-forming gas.

5) At $M_\ast>10^{10}\msun$, the SF galaxies in clusters show much
 lower typical specific SFRs than galaxies in groups and the field,
 due to a more significant population of massive galaxies with very low SFR
 (14\% of all the cluster galaxies with $M_\ast>10^{10}\msun$).  This could result from ram pressure
 stripping being less efficient in removing gas from more massive
 cluster galaxies, allowing such galaxies to sustain low-level star
 formation fueled by a residual gas reservoir.

\acknowledgments

We thank Augustus Oemler, Alan Dressler, and Ann Zabludoff for useful
discussions, and the referee for insightful comments and
suggestions. JR acknowledges support provided by the National
Aeronautics and Space Administration through Chandra Postdoctoral
Fellowship Award Number PF7-80050 issued by the Chandra X-ray
Observatory Center, which is operated by the Smithsonian Astrophysical
Observatory for and on behalf of the National Aeronautics and Space
Administration under contract NAS8-03060. We also acknowledges partial
support for this work from NASA-JPL grant 1310258, NASA grant
NNG04GC846 and NSF grant AST-0707417.  This publication makes use of
data products from the Two Micron All Sky Survey, which is a joint
project of the University of Massachusetts and the Infrared Processing
and Analysis Center/California Institute of Technology, funded by the
National Aeronautics and Space Administration and the National Science
Foundation.

\clearpage
\bibliographystyle{apj}

\clearpage

\begin{deluxetable}{lcccccc}
\tablecaption{Individual group properties.}
\tablehead{Group&RA\tablenotemark{a} (J2000)&Dec\tablenotemark{a} (J2000)& N\tablenotemark{b} &N$_{IMACS}$\tablenotemark{c}  & z\tablenotemark{d} & $\sigma$\tablenotemark{e} (km s$^{-1}$)}

\startdata
MZ\,10451&02:29:13.17&  -29:38:58.1& 60&34 & 0.06065   &   $503_{-71}^{+56}$\\
MZ\,3849 &10:27:49.26&  -03:18:25.4& 20&15 & 0.06054   &   $298_{-42}^{+30}$\\
MZ\,4577 &11:32:43.87&  -03:57:55.7& 35&17 & 0.06201   &   $247_{-41}^{+35}$\\
MZ\,4592 &11:30:46.48&  -03:47:56.2& 27&16 & 0.06162   &   $215_{-36}^{+23}$\\
MZ\,4940 &11:35:58.66&  -03:41:05.0& 12&8  & 0.06212   &   $104_{-58}^{+31}$\\
MZ\,5293 &12:16:25.78&  -03:24:25.9& 10&9  & 0.06204   &    $99_{-24}^{+11}$\\
MZ\,5383 &12:35:01.07&  -03:36:11.3& 47&25 & 0.06044   &   $521_{-63}^{+54}$\\
MZ\,9014 &00:38:05.40&  -27:23:53.9& 34&23 & 0.06094   &   $322_{-40}^{+30}$\\
MZ\,9307 &00:40:20.57&  -27:32:17.7& 28&16 & 0.05999   &   $431_{-72}^{+44}$\\
\enddata
\tablenotetext{a}{Coordinates of the luminosity-weighted group centers.}
\tablenotetext{b}{Total number of group members.}
\tablenotetext{c}{Number of group members with new redshifts measured by IMACS.}
\tablenotetext{d}{Biweight mean redshift of group members.}
\tablenotetext{e}{Velocity dispersion.}
\label{tab_1}
\end{deluxetable}

\end{document}